\documentclass[11pt]{article}

\usepackage[top=60pt,bottom=70pt,left=75pt,right=75pt]{geometry}

\usepackage{amsmath, amssymb, amsfonts, latexsym, euscript, mathalfa, mathrsfs}
\usepackage{graphicx, color, epsfig, xcolor}
\usepackage{cite}
\usepackage{empheq}

\usepackage{setspace, sectsty}

\usepackage{braket, cancel}

\usepackage[debug,pageanchor=false]{hyperref}
\definecolor{link}{rgb}{.8,.15,.1}
\hypersetup{colorlinks=true,linkcolor=link,citecolor=link,urlcolor=link,linktocpage}

\makeatletter
\@addtoreset{equation}{section}
\makeatother

\newlength{\sswidth}

\newcommand{\nn}{\nonumber}

\newcommand{\R}{\mathrm{Re}}    
\newcommand{\I}{\mathrm{Im}}       
\newcommand{\vol}{\mathrm{vol}}

\newcommand{\g}{\gamma}

\newcommand{\e}{\mathrm{e}}

\newcommand{\phpa}{\phi^{++}_+ + \phi^{--}_+}

\newcommand{\phpd}{\phi^{++}_+ - \phi^{--}_+}

\newcommand{\phmb}{\phi^{+-}_- + \phi^{-+}_-}
\newcommand{\phmc}{\phi^{+-}_- - \phi^{-+}_-}

\begin{document}

\begin{titlepage}

\begin{flushright} \small
UUITP-32/17
\end{flushright}

\begin{center}

\noindent

{\Large \bf $\mathcal{N} = 2$ supersymmetric AdS$_4$ solutions \\ \vspace{.3cm} of type IIB supergravity}

\bigskip\medskip

Achilleas Passias$^1$, Gautier Solard$^2$ and Alessandro Tomasiello$^2$\\

\bigskip\medskip
{\small 

$^1$Department of Physics and Astronomy, Uppsala University,\\
Box 516, SE-75120 Uppsala, Sweden
\\	
\vspace{.3cm}
$^2$Dipartimento di Fisica, Universit\`a di Milano--Bicocca, \\ Piazza della Scienza 3, I-20126 Milano, Italy \\ and \\ INFN, sezione di Milano--Bicocca
	
}

\vskip .5cm 
{\small \tt achilleas.passias@physics.uu.se, gautier.solard@mib.infn.it, alessandro.tomasiello@unimib.it}

\vskip .9cm 
     	{\bf Abstract }
\vskip .1in
\end{center}

\noindent
We analyze general $\mathcal{N}=2$ supersymmetric AdS$_4$ solutions of
type IIB supergravity.
Utilizing a set of pure spinor equations directly adapted to $\mathcal{N}=2$, 
the necessary and sufficient conditions for supersymmetry are reduced to a 
concise system of partial differential equations for two functions which determine the solutions. 
We show that using this system analytic solutions can be generated, thus potentially expanding
the rather limited set of known AdS$_4$ solutions in type IIB supergravity.

\noindent

\vfill
\eject

\end{titlepage}

\tableofcontents

\section{Introduction}

Compactifications of string theories with a negative cosmological constant, although not realistic cosmologically, are
more abundant than those with a positive one, and in the framework of the AdS/CFT correspondence serve as a description of strongly coupled conformal field theories (CFTs); \emph{vice versa}, CFT intuition can often suggest new classes of anti-deSitter (AdS) solutions. 

In the context of AdS$_4$/CFT$_3$ correspondence, solutions with extended supersymmetry (${\cal N}\ge 2$) are far easier to deal with than ${\cal N}=1$ solutions. The latter are dual to CFT$_3$'s with only two Poincar\'{e} or $Q$ supercharges, which usually do not provide much computational power.

A prominent class of AdS$_4$ solutions with ${\cal N}= 2$ supersymmetry is the one of compactifications of M-theory on Sasaki--Einstein seven-manifolds (Freund--Rubin class), arising as the near-horizon geometries of M2-branes probing Calabi-Yau four-fold singularities. They can be generalized by additional flux on the internal space (see \cite{Gabella:2012rc} for a general analysis), or reduced to type IIA supergravity (see e.g.~\cite{martelli-sparks-sasaki7} for an explicit discussion) where can also be modified by adding fluxes \cite{gaiotto-t2,petrini-zaffaroni,lust-tsimpis-singlet-2,ajtz}. In the first case, \cite{Gabella:2012rc}, $\mathcal{N}=2$ supersymmetry was imposed from the outset, whereas some solutions in the latter case relied on ${\cal N}=1$ classifications \cite{lust-tsimpis, gmpt3}, ensuring enhanced supersymmetry by 
imposing the presence of a U$(1)$ symmetry corresponding to the U$(1)$ R-symmetry\footnote{Namely an isometry which does not leave the supercharges invariant, and can thus be used to generate more of them.} of the ${\cal N} = 2$ superalgebra. This is in turn was achieved using an Ansatz inspired by the reduction of the Sasaki--Einstein solutions \cite{petrini-zaffaroni}. More recently, ${\cal N} = 2$ AdS$_4$ solutions were found in massive type IIA supergravity 
by uplifting solutions of a four-dimensional gauged supergravity \cite{Guarino:2015jca}.

In this paper, we impose ${\cal N}=2$ supersymmetry in full generality, using an extension of the pure spinor approach \cite{gmpt2, gmpt3}. In the ${\cal N}=1$ case, the internal part of the supercharges defines a pair of polyforms $\phi_\pm$ on the internal manifold $M_6$, called pure spinors, which satisfy a system of differential equations \cite{gmpt3}. In the ${\cal N}=2$ case, one can define a $2\times2$ matrix $\phi_\pm^{IJ}$ of such pure spinors. As described above, one can impose the differential equations on one entry of this matrix, say $\phi_\pm^{11}$, and rely on R-symmetry to generate the others. Here we choose instead to derive a system of ``extended pure spinor equations'' on $\phi_\pm^{IJ}$ which ensures directly ${\cal N}=2$ supersymmetry. R-symmetry is then obtained as a by-product.
 
Deriving the extended pure spinor equations can be done relatively painlessly by using the ten-dimensional approach \cite{10d}. This consists of a system that can be applied to any supersymmetric solution (even with one supercharge), without an AdS$_4\times M_6$ or any other factorization. It was checked in \cite{10d} that it reproduces quickly and correctly the pure spinor equations \cite{gmpt2}; here we use it in a similar fashion to deal with extended supersymmetry. Some of the equations in the system we obtain are a natural extension to the whole matrix $\phi_\pm^{IJ}$ of the ${\cal N}=1$ system that one would apply to one entry; others are new. Most notably, one of the equations gives an expression for the Ramond--Ramond (R--R) fluxes that does not involve the Hodge star. From this it follows that all the Bianchi identities for the R--R fluxes are automatically satisfied, which is typically not the case for systems with ${\cal N}=1$ supersymmetry. 

We decided to apply this new system to type IIB supergravity, where supersymmetric AdS$_4$ solutions appear to be scarcer than in type IIA: with minimal supersymmetry there are a few isolated examples \cite{lust-tsimpis-singlet-2,solard-IIB}, while with extended supersymmetry there exists a notable class \cite{assel-bachas-estes-gomis}, based on work in \cite{dhoker-estes-gutperle}, which has ${\cal N}=4$ supersymmetry and is the dual of the Hanany--Witten theories \cite{hanany-witten}. Other solutions with extended supersymmetry have been obtained by applying non-abelian T-duality transformations on type IIA solutions \cite{Lozano:2016wrs, PandoZayas:2017ier}, or uplifting lower-dimensional vacua \cite{Inverso:2016eet}.

After parameterizing the pure spinors in terms of an identity structure on the internal manifold $M_6$, and ``running'' the extended pure spinor equations, we obtain a set of differential equations for the identity structure. (This in turn gives rise to an SU(3) structure closely resembling the structure of the aforementioned Ansatz usually employed in type IIA \cite{petrini-zaffaroni,lust-tsimpis-singlet-2,ajtz}.) As it often happens, many of the equations are redundant, and in the end only a small set of rather simple equations survives, which can be interpreted as defining local coordinates. One in particular defines a transversely-holomorphic foliation.\footnote{A similar foliation also appeared for example in the study of supersymmetric theories on curved spaces \cite{ktz,Closset:2012ru}.} 

Using these local coordinates, we can finally reduce the entire system to three partial differential equations (PDEs) for two functions. They are relatively simple in form, and evoke results obtained for other similar problems. One of the equations, for example, is a version with a source term of the Toda equation that appeared in \cite{lin-lunin-maldacena}. 

Exploring the space of solutions to this system is an elaborate task which we will not undertake here. We do however describe a couple of elementary Ans\"atze that simplify the system, so that we recover the maximally supersymmetric AdS$_5\times S^5$ solution (considered as a warped AdS$_4$ solution) and generate a few new formal solutions. While we are not certain that there is a compact and physical $M_6$ among these, further study of the PDEs is likely to be rewarding.

In section \ref{sec:reduction} we will describe how to obtain our extended pure spinor equations from the ten-dimensional system of \cite{10d}. After introducing a parameterization for the pure spinors $\phi_\pm$ in section \ref{sec:param}, we will analyze the equations in section \ref{sec:analysis}, obtaining a relatively simple set of conditions summarized in section \ref{sub:sum1}. As is often the case, these conditions will suggest a choice of local coordinates, which we will use in section \ref{sec:local} to simplify the equations further, arriving at our final system in section \ref{total}. We will end in section \ref{sec:sol} by discussing a few solutions.

\section{Reduction of the $10d$ supersymmetry equations}
\label{sec:reduction}

In \cite{10d} a system of equations was obtained, which constitute necessary and sufficient conditions for any ten-dimensional solution of type II supergravity to preserve superymmetry. We will specialize this system to the case of an AdS$_4$ background of type IIB supergravity, preserving $\mathcal{N}=2$ supersymmetry.

Let us review the system of equations of \cite{10d}, which are summarized in section 3.1 of that paper.
Let us focus on the following subset of equations:
\begin{subequations}
\begin{align}
d_H(e^{-\phi} \Phi) &= -(\widetilde K \wedge + \iota_K) F_{(10d)} \ , \label{10d-eq1} \\
d \widetilde K &= \iota_K H \ . \label{10d-eq2}
\end{align}
\end{subequations} 
Here $\phi$ is the dilaton, $H$ is the NS--NS three-form field strength, $d_H \equiv d - H\wedge$, and $F_{(10d)}$ is the sum of the R--R field strengths. The latter sum, following the ``democratic formulation'' of type II supergravities, includes all the $p$-form field strengths, with $p$ odd for type IIB, subject to the self-duality constraint $F=*\lambda(F)$. $\lambda$ is an operator acting on a $p$-form $F_p$ as $\lambda(F_p) = (-1)^{\left[p/2\right]} F_p$ where square brackets
denote the integer part.

$\Phi$ is a bispinor constructed out of the supersymmetry parameters $\epsilon_1$ and $\epsilon_2$:
\begin{equation}
\Phi \equiv \epsilon_1 \overline{\epsilon_2} \ .
\end{equation} 
The latter are Majorana--Weyl spinors of positive chirality. $K$ and $\widetilde K$ are respectively a vector and a 1-form
bilinear:
\begin{equation}
K \equiv \tfrac{1}{64}(\overline{\epsilon_1} \Gamma^M \epsilon_1 + \overline{\epsilon_2} \Gamma^M \epsilon_2) \partial_M \ , \qquad
\widetilde K \equiv \tfrac{1}{64}(\overline{\epsilon_1} \Gamma_M \epsilon_1 - \overline{\epsilon_2} \Gamma_M \epsilon_2) dx^M \ ,
\end{equation}
with $K$ being a Killing vector, and more general the generator of a symmetry of the full solution.

We now turn to applying these equations to the AdS$_4$ background of interest. To do so we will make a ``$4+6$'' split of the ten-dimensional fields and the supersymmetry parameters.

We want to allow for the most general geometry with an AdS$_4$ factor, leaving the symmetries of the latter intact. This amounts to taking the $10d$ spacetime to be a warped product of AdS$_4$ and a six-dimensional manifold $M_6$, with the warp factor being a function only on $M_6$. The corresponding line element is:
\begin{equation}
ds^2_{10} = e^{2A} ds^2_{\mathrm{AdS}_4} + ds^2_{M_6} \ , 
\end{equation} 
where $A$ is the warp factor.

Accordingly, the $H$ field is a form only on $M_6$, while the R--R field strengths are decomposed as
\begin{equation}
F_{(10d)} = e^{4A} \vol_4 \wedge * \lambda(F) + F \ , 
\qquad F = F_1 + F_3 + F_5 \ .
\end{equation} 

Turning to the supersymmetry parameters, we will take them to be a product of Spin$(1,3)$ and Spin$(6)$ spinors. For an $\mathcal{N}=1$ supersymmetric AdS$_4$ solution this decomposition is:
\begin{equation}
\epsilon_i = \chi_+ \otimes \eta_{i+} + \chi_- \otimes \eta_{i-}, \qquad i=1,2 \ ,
\end{equation}
where the $\chi$'s are AdS$_4$ Killing spinors and the $\eta$'s spinors on $M_6$. A plus or minus subscript denotes the chirality of the spinor. 
Since we are interested in $\mathcal{N}=2$ supersymmetry we need to add a second pair of $\chi$'s and $\eta$'s. The decomposition Ansatz thus becomes
\begin{equation}\label{spin_decomp}
\epsilon_i = \sum_{I = 1}^2 \chi_+^I \otimes \eta_{i+}^I + \sum_{J = 1}^2 \chi_-^J \otimes \eta_{i-}^J \ ,
\end{equation}
with $I,\, J$ indices upon which an SO$(2)$ R-symmetry acts. 

As noted, the $\chi$'s are AdS$_4$ Killing spinors, i.e.~they satisfy\footnote{Using 
\begin{equation}\label{Killing}
\chi_-^I = B (\chi_+^I)^* \ , \qquad
\g_\mu B =  B \g^*_\mu \ , \qquad
\g^0 \g_\mu = - \g_\mu^\dagger \g^0 \ .
\end{equation}
See appendix \ref{spin(1,3)} for more details on Cliff$(1,3)$ conventions.
}
\begin{equation}
\nabla_\mu \chi^I_\pm = \frac{1}{2} \gamma_\mu \chi^I_\mp \ , \qquad
\nabla_\mu \overline{\chi^I_\pm} = - \frac{1}{2} \overline{\chi^I_\mp} \gamma_\mu \ .
\end{equation}
We will consider the case that $\chi^1_+$ and $\chi^2_+$ are linearly independent, since otherwise we would only have $\mathcal{N} = 1$ supersymmetry. To see this consider $\chi^2_+ = a \chi^1_+$; since both $\chi^1_+$ and $\chi^2_+$ satisfy the Killing spinor equation it is easy to see that in fact $a$ is a constant. Then
\begin{equation}
\epsilon_i = \chi_+^1 \otimes (\eta^1_{i+} + a \eta^2_{i+}) + {\rm c.c.} \ ,
\end{equation} 
where c.c. denotes the complex conjugates. Since $a$ is constant we can define $\tilde \eta_i = \eta^1_{i+} + a \eta^2_{i+}$ and we end up with an $\mathcal{N} = 1$ decomposition.

Finally, in reducing the 10$d$ equations we will use the following decomposition of Cliff(1,9):
\begin{equation}\label{cliff_decomp}
\Gamma_\mu = e^A \gamma^{(4)}_\mu \otimes \mathbb{I} \ , \qquad
\Gamma_{m + 3} = \g_5^{(4)} \otimes \gamma^{(6)}_m \ , \qquad
\Gamma_{11} \equiv \Gamma^0 \dots \Gamma^{9} = \gamma^{(4)}_5 \otimes \gamma^{(6)}_7 \ ,
\end{equation}
with $\mu = 0,1,2,3$ and $m=1,2,\dots 6$. $\g_5^{(4)}$ and $\gamma^{(6)}_7$ are the chirality operators in $1+3$ and $6$
dimensions respectively.

We can now proceed with the reduction.

We first look at \eqref{10d-eq2}. $\widetilde{K}$, $K$ decompose as\footnote{Henceforth, we drop the $(4)$ and $(6)$ superscripts from the gamma matrices.} 
\begin{subequations}
\begin{align}
\widetilde K _\mu &= \frac{1}{32}\sum_{I,J=1}^{2} \overline{\chi^I_+} \gamma_\mu \chi^J_+ e^A (\overline{\eta^I_{1+}} \eta^J_{1+} - \overline{\eta^I_{2+}} \eta^J_{2+}) \ , \\
\widetilde K_m &= -\frac{1}{16} \R ( \overline{\chi^1_+} \chi^2_- \tilde\xi_m ) \ , \\
K^\mu &= \frac{1}{32}\sum_{I,J=1}^{2} \overline{\chi^I_+} \gamma^\mu \chi^J_+ e^{-A} (\overline{\eta^I_{1+}} \eta^J_{1+} + \overline{\eta^I_{2+}} \eta^J_{2+}) \ , \\
K^m &= -\frac{1}{16} \R ( \overline{\chi^1_+} \chi^2_- \xi^m ) \ ,
\end{align}
\end{subequations}
where
\begin{equation}
\tilde\xi_m \equiv \overline{\eta^1_{1+}} \gamma_m \eta^2_{1-} - \overline{\eta^1_{2+}} \gamma_m \eta^2_{2-} \ , \qquad \xi^m \equiv \overline{\eta^1_{1+}} \gamma^m \eta^2_{1-} + \overline{\eta^1_{2+}} \gamma^m \eta^2_{2-} \ .
\end{equation}
We thus find
\begin{equation}\label{eq:dfdxi}
\overline{\eta_{1+}^{(I}} \eta_{1+}^{J)} = 
\overline{\eta_{2+}^{(I}} \eta_{2+}^{J)} \ , \qquad
d(e^A f)  =  -\frac{1}{2} \I(\tilde \xi)  \ , \qquad 
d \tilde \xi = i_\xi H \ , 
\end{equation}
where
\begin{equation}
2 \epsilon^{IJ} f \equiv -i\overline{\eta_{1+}^{[I}} \eta_{1+}^{J]} + i\overline{\eta_{2+}^{[I}} \eta_{2+}^{J]} \ ,  
\end{equation}
$\epsilon^{IJ}$ being the Levi--Civita symbol with $\epsilon^{12} = 1$.

Next, we impose the condition that $K$ is a Killling vector i.e.\ $\nabla_{(M} K_{N)} = 0$. Doing so we get 
\begin{equation}\label{spinor_scalars}
\overline{\eta_{1+}^{(I}} \eta_{1+}^{J)} = 
\overline{\eta_{2+}^{(I}} \eta_{2+}^{J)} \equiv \frac{1}{2} c^{IJ} e^A \ , \qquad
-i\overline{\eta_{1+}^{[I}} \eta_{1+}^{J]} = i\overline{\eta_{2+}^{[I}} \eta_{2+}^{J]} \equiv \epsilon^{IJ} f \ ,  
\end{equation}
where $c^{IJ}$ are constants. 
In addition
\begin{equation}
\I (\xi) = 0 \ , \qquad \nabla_{(n} \xi_{m)} = 0 \ ;
\end{equation}
thus \emph{$\xi$ is a Killing vector}.

Next comes equation \eqref{10d-eq1}. In order to reduce \eqref{10d-eq1}, we need to write $\Phi$ as a product of external and internal (poly)forms. To do so we decompose the Fierz expansion of $\Phi$, utilizing \eqref{spin_decomp} and \eqref{cliff_decomp}. We find\footnote{Powers of $e^A$ coming from \eqref{cliff_decomp} have been suppressed.}
\begin{equation}
\Phi = \sum_{IJ} (
\chi^I_+\overline{\chi^J_+} \wedge \eta^I_{1+} \overline{\eta^J_{2+}} + 
\chi^I_+\overline{\chi^J_-} \wedge \eta^I_{1+} \overline{\eta^J_{2-}} -
\chi^I_-\overline{\chi^J_+} \wedge \eta^I_{1-} \overline{\eta^J_{2+}} +
\chi^I_-\overline{\chi^J_-} \wedge \eta^I_{1-} \overline{\eta^J_{2-}}
) \ .
\end{equation}
From \eqref{10d-eq1} we want to obtain differential conditions for the ``internal'' bispinors and in order to do so we need
the derivatives of the ``external'' ones. The latter can be derived from \eqref{Killing}:
\begin{subequations}
\begin{align}
d(\chi_{\pm}^{I}\overline{\chi_{\pm}^{J}}) 
&= 2 \left(1-\tfrac{1}{4}(-1)^k(4-2k)\right) \R (\chi_{\mp}^{I}\overline{\chi_{\pm}^{J}}) \ , \\
d(\chi_{\pm}^{I}\overline{\chi_{\mp}^{J}}) 
&= 2i\left(1+\tfrac{1}{4}(-1)^k(4-2k)\right) \I (\chi_{\mp}^{I}\overline{\chi_{\mp}^{J}}) \ ,
\end{align}
\end{subequations}
where $k$ is the degree of the individual components of the bispinor, considered as a polyform.

Schematically, \eqref{10d-eq1} then becomes
\begin{equation}
{\rm ext} \wedge \left[ {\rm int} + d_{H}\left({\rm int}\right) \right] = F \ ,
\end{equation}
where ``${\rm ext}$'' represents collectively the external part of $\Phi$, ``${\rm int}$'' the internal part,
and $F$ the term involving the R--R fluxes. Next, the external part is expanded in linearly independent $p$-form components. Each resulting term has to vanish separately, giving an equation for the internal part. More details can be found in appendix A of \cite{afprt}. The calculation there is for AdS$_6 \times M_4$ backgrounds but the procedure is essentially the same. 

In the end we obtain  
\begin{subequations}\label{eq:pureIJ}
\begin{align}
d_H\left(e^{2A-\phi} \phi^{(IJ)}_-\right) + 2 e^{A-\phi} \R \phi_+^{(IJ)} &= 0 \ , \label{eq:pureIJ1} \\ 
d_H\left(e^{3A - \phi} \R \phi_+^{[IJ]}\right) &= 0 \ , \label{eq:pureIJ2}\\
d_H\left(e^{A-\phi} \I \phi^{[IJ]}_+\right) + e^{-\phi} \I \phi_-^{[IJ]} &= -\frac{1}{8} e^A f F \epsilon^{IJ}\ ; 
\label{RRflux} 
\end{align}
and
\begin{align}
d_H\left(e^{3A-\phi} \I \phi_+^{(IJ)}\right) + 3 e^{2A-\phi} \I \phi_-^{(IJ)} &= -\frac{1}{16} c^{IJ} e^{4A} * \lambda(F) \ , \label{eq:pureIJ4} \\
d_H\left(e^{-\phi} \phi_-^{[IJ]}\right) &= -\frac{1}{16}(\bar{\tilde \xi} \wedge + \iota_\xi) F \epsilon^{IJ} \ , \label{eq:pureIJ5} \\
d_H\left(e^{4A-\phi} \phi_-^{[IJ]}\right) + 4 e^{3A-\phi}  \R \phi_+^{[IJ]}&= -\frac{i}{16}(\bar{\tilde \xi} \wedge + \iota_\xi) e^{4A} * \lambda(F)  \epsilon^{IJ}\ ,
\label{eq:pureIJ6}
\end{align}
\end{subequations}
where
\begin{equation}
\phi_+^{IJ} \equiv \eta^I_{1+}\overline{\eta^J_{2+}} 
\ , \qquad 
\phi_-^{IJ} \equiv \eta^I_{1+}\overline{\eta^J_{2-}} \ ,
\end{equation}
and $\bar{\tilde \xi}$ is the complex conjugate of $\tilde{\xi}$.

Although the system (\ref{eq:pureIJ}) appears large, in fact it has a high degree of redundancy: for instance (and as we will see in the sections that follow) $c^{IJ}$ can be set proportional to the identity, and
following that, except for \eqref{RRflux}, the equations that involve the R--R fields are redundant. (This is why we have separated the equations in two blocks.)

The system (\ref{eq:pureIJ}) is also redundant in another, more trivial way. Consider its diagonal components, $I=J$. Then only the two equations (\ref{eq:pureIJ1}), (\ref{eq:pureIJ4}) survive: they are two copies of the pure spinor equations \cite{gmpt3} for ${\cal N}=1$ AdS$_4$ solutions. Solving them gives by definition two solutions of the supersymmetry equations, with the same fluxes and geometry; in other words, it gives an ${\cal N}=2$ solution. Thus the $I\neq J$ equations are redundant. 

Even though (\ref{eq:pureIJ}) is highly redundant, it will be more convenient for our analysis. For example, some of the information that would appear at high form order in the subsystem (\ref{eq:pureIJ1}), (\ref{eq:pureIJ2}), (\ref{RRflux}) appears at lower form order in the full system (\ref{eq:pureIJ}), and is easier to handle. 

We can now also comment about the remaining equations in \cite{10d}, called (3.1c) and (3.1d). Those ``pairing equations'' are in general needed, but for AdS$_4$ vacua they are redundant. Indeed, as we have remarked, the ${\cal N}=1$ supersymmetry system is already reproduced by (\ref{eq:pureIJ1}), (\ref{eq:pureIJ4}) above. (How exactly they become redundant was shown in \cite[Sec.~4]{10d} for Minkowski$_4$; that logic can be adapted to AdS$_4$ once again following \cite[App.~A]{afprt}.) Thus, the pairing equations are not needed for our ${\cal N}=2$ classification; they would make our system (\ref{eq:pureIJ}) even more redundant. We are free to ignore them, and in the following we have done so.

\section{Parametrization of the pure spinors}
\label{sec:param}
The spinors $\eta_{i+}^I$ define an identity structure in six dimensions; see appendix \ref{spin(6)}. In this section we will introduce a set of 1-forms parametrizing the latter and express the pure spinors $\phi_\pm^{IJ}$ in terms of these. Before doing so we will manipulate the results of the previous section in two ways.

The first one is fixing the constants $c^{IJ}$ of \eqref{spinor_scalars} as
\begin{equation}\label{eq:cIJd}
c^{IJ} = 2 \delta^{IJ} \ ,
\end{equation} 
where $\delta^{IJ}$ is the Kronecker delta. We can do so because the decomposition Ansatz \eqref{spin_decomp} doesn't fix the spinors $\eta^I_{i+}$ uniquely. 
Specifically, one is free to make a GL$(2,\mathbb{R})$ transformation that leaves the fixed (by \eqref{spinor_scalars}) norms
$\|\eta_{i+}^I\| = e^A$ invariant, leading to real linear combinations of the external spinors $\chi_+^I$. The details of this transformation can be found in appendix \ref{GL2R}. 
Note that since $c^{12} = \overline{\eta_{i+}^{(1}} \eta_{i+}^{2)} = 0$, from $\overline{\eta_{i+}^{[1}} \eta_{i+}^{2]} = \overline{\eta_{i+}^{1}} \eta_{i+}^{2}$ and $|\overline{\eta^1_{i+}}\eta^2_{i+}| \leq \sqrt{\|\eta^1_{i+}\|\|\eta^2_{i+}\|}$ it follows that
\begin{equation}\label{f_ineq}
|f| \leq e^A \ .
\end{equation}

The second one is that instead of $\eta^I_{i+}$ we will work with
\begin{equation}
\eta^\pm_{i+} = \frac{1}{\sqrt{2}} (\eta^1_{i+} \pm i \eta^2_{i+})
\end{equation}
which have charge $\pm 1$ under the U$(1)\simeq$ SO$(2)$ R-symmetry.
The conditions \eqref{spinor_scalars} (with $c^{IJ} = 2 \delta^{IJ}$) become
\begin{equation}\label{spinor_scalarsII}
\overline{\eta^\pm_{i+}} \eta^\mp_{i+} = 0 \ , \qquad
\overline{\eta^\pm_{1+}} \eta^\pm_{1+} = f_\mp \ , \qquad
\overline{\eta^\pm_{2+}} \eta^\pm_{2+} = f_\pm \ , 
\end{equation}
where $f_\pm \equiv e^A \pm f$.

Given a chiral spinor $\eta_+$ of positive chirality (and its complex conjugate $\eta_- \equiv (\eta_+)^c)$, we can express $\eta_{i+}^\pm$, taking into account \eqref{spinor_scalarsII}, as follows:
\begin{subequations}
\begin{align}
\eta^+_{1+} &= \sqrt{f_-} \eta_+ \ , \\
\eta^-_{1+} &= \sqrt{f_+} \frac{1}{2} w_1 \eta_- \ , \\
\eta^+_{2+} &= \sqrt{f_+} \left(a \eta_+ + \frac{1}{2} b w_3 \eta_-\right) \ , \\
\eta^-_{2+} &= \sqrt{f_-} \frac{1}{2} c w_2 \left(a^* \eta_- - \frac{1}{2} b \overline{w_3}\eta_+\right) \ .
\end{align}
\end{subequations}
Here $a \in \mathbb{C}$ and $b, c \in \mathbb{R}$. They satisfy
\begin{align}\label{abc}
|a|^2 + b^2 = 1 \ ,  \qquad
c^{-1} = \left(|z_1|^2b^2+|a|^2\right)^{1/2} \ ,
\end{align}
with $z_1$ defined below. The 1-forms $\{ w_1, w_2, w_3 \}$ parametrize the identity structure and are holomorphic with respect to the almost complex structure $J$ defined by $\eta_+$; see appendix \ref{spin(6)}.

We introduce
\begin{equation}
z_1 \equiv \frac{1}{2}(w_2,w_3) \ , \qquad 
z_2 \equiv \frac{1}{2}(w_3,w_1) \ , \qquad 
z_3 \equiv \frac{1}{2}(w_1,w_2) \ ,
\end{equation} 
where $(\cdot,\cdot)$ denotes the inner product. 
We then have
\begin{equation}
(w_a,w_b) = 2 Z_{ab} \ , \qquad
Z \equiv
\begin{pmatrix}
1 & z_3 & z^*_2 \\
z_3^* & 1 & z_1 \\
z_2 & z_1^* & 1 
\end{pmatrix} \ .
\end{equation}
The determinant of $Z$ is 
\begin{equation}
\det Z = 1-|z_1|^2 - |z_2|^2 - |z_3|^2 - 2 \R(z_1z_2z_3) \ .
\end{equation}

The pair $(J, \Omega)$ that characterize the SU$(3)$ structure defined by $\eta_+$ are expressed in terms of $\{ w_1, w_2, w_3 \}$ as
\begin{equation}\label{eq:JO}
J = \frac{i}{2} (Z^{-1})^{ab} w_a \overline{w_b} \ , \qquad \Omega = \frac{e^{i \vartheta}}{\sqrt{\det Z}} w_1 \wedge w_2 \wedge w_3 \ 
\end{equation}
where $\vartheta \in \mathbb{R}$.

We can now express the pure spinors 
\begin{equation}
\phi_+^{\pm\pm} \equiv \eta^\pm_{1+}\overline{\eta^\pm_{2+}} 
\ , \qquad 
\phi_-^{\pm\pm} \equiv \eta^\pm_{1+}\overline{\eta^\pm_{2-}} \ ,
\end{equation}
in terms of forms: 
\begin{subequations}
\begin{align}
\phi^{++}_+ &= \frac{1}{8}\sqrt{f_+f_-}\left[a^* e^{-iJ} + \frac{1}{2} b (\overline{w_3}\wedge \Omega-\Omega_{\overline{w_3}})\right] \ , \\
\phi^{++}_- &=  \frac{1}{8}\sqrt{f_+f_-}\left[- a \Omega - b w_3 \wedge e^{-iJ} \right] \ , \\
\phi^{+-}_+ &= \frac{1}{8}f_- \left[\frac{1}{2} ac (\overline{w_2}\wedge \Omega-\Omega_{\overline{w_2}}) -b c z_1  e^{-iJ} \right] \ , \\ 
\phi^{+-}_- &= \frac{1}{8}f_- \left[-a^* c w_2 \wedge e^{-iJ} +b c z_1^* \Omega\right] \ , \\ 
\phi^{-+}_+ &= \frac{1}{8}f_+ \left[\frac{1}{2}a^*(w_1\wedge \overline{\Omega}+\overline{\Omega}_{w_1}) + b z_2 e^{iJ}+b w_1\wedge \overline{w_3}\wedge e^{iJ} \right] \ , \\
\phi^{-+}_- &= \frac{1}{8}f_+ \left[a w_1 \wedge e^{iJ} - \frac{1}{4} b (w_1,w_3,\overline{\Omega})\right] \ , \\
\phi^{--}_+ &= \frac{1}{8} \sqrt{f_+f_-} \left[a c z_3^* e^{iJ} +a c w_1\wedge \overline{w_2}\wedge e^{iJ}- \frac{1}{2} b c z_1 (w_1\wedge \overline{\Omega}+\overline{\Omega}_{w_1})\right] \ , \\
\phi^{--}_- &= \frac{1}{8}\sqrt{f_+f_-} \left[- \frac{1}{4} a^*c (w_1,w_2,\overline{\Omega}) - bcz_1^* w_1 \wedge e^{iJ}\right] \ .
\end{align}
\end{subequations}
In the above
\begin{equation}
(u,w,\overline{\Omega}) \equiv \iota_u\iota_w \overline{\Omega}+ u \wedge \iota_w \overline{\Omega} + w \wedge \iota_u \overline{\Omega} - u \wedge w \wedge \overline{\Omega} \ .
\end{equation}

We also have
\begin{subequations}
\begin{align}
(\xi)^\flat &= i\sqrt{f_+f_-}\left[\overline{w_1}+b^2c z_1\overline{w_3}+ |a|^2c\overline{w_2}-\frac{1}{4}abc\, \iota_{\overline{w_2}}\iota_{\overline{w_3}}\Omega\right] \ , \\
\tilde{\xi} &= i\sqrt{f_+f_-}\left[\overline{w_1}-b^2c z_1\overline{w_3}- |a|^2c\overline{w_2}+\frac{1}{4}abc\,\iota_{\overline{w_2}}\iota_{\overline{w_3}}\Omega\right] \ ,
\end{align}
\end{subequations}
where $(\xi)^\flat$ is the 1-form dual to the $\xi$ vector.

\section{Analysis of the supersymmetry equations}
\label{sec:analysis}

In this section we initiate the analysis of the supersymmetry equations obtained in section \ref{sec:reduction}. 
We will first analyze those which do not involve the R--R field strengths, leaving the analysis of the latter for the end.
As we anticipated, not all the equations are independent, and we will be able to reduce them to a significantly smaller set.

\subsection{System of equations}
After switching from the $\phi^{IJ}_\pm$ to the $\phi_\pm^{\pm\pm}$ pure spinors introduced in the previous section,
the system of supersymmetry equations is as follows:
\begin{subequations}\label{SUSYeq}
\begin{align}
d_H\left[e^{2A-\phi} \phi^{++}_-\right]+e^{A-\phi}(\phi^{+-}_+ + \overline{\phi^{-+}_+}) &= 0 \ , \label{S1} \\
d_H\left[e^{2A-\phi}(\phmb)\right]+2e^{A-\phi}\R(\phpa) &= 0 \ , \label{S2} \\
d_H\left[e^{2A-\phi} \phi^{--}_-\right]+e^{A-\phi}(\overline{\phi^{+-}_+}+ \phi^{-+}_+)&= 0 \ , \label{S3} \\
d_H\left[e^{3A-\phi}\I(\phpd)\right] &= 0 \ , \label{S4} \\
d_H\left[e^{3A-\phi}(\phi^{+-}_+ -\overline{\phi^{-+}_+})\right]+3e^{2A-\phi}(\phi^{++}_- - \overline{	\phi^{--}_-}) &= 0 \label{S5} \ , \\ 
d_H\left[e^{A-\phi}\R(\phpd)\right]+e^{-\phi}\R(\phmc) &= - \frac{1}{4}e^{A} f F \ , \label{F2} 
\end{align}
\end{subequations}
and
\begin{subequations}
\begin{align}
d_H\left[e^{3A-\phi}\I(\phpa)\right]+3e^{2A-\phi}\I(\phmb) &= -\frac{1}{4}e^{4A}\ast\lambda(F) \ ,  \label{F1} \\ 
d_H\left[e^{-\phi}(\phmc)\right] &= \frac{i}{8}(\bar{\tilde{\xi}}\wedge+\iota_{\xi})F \ , \label{F3} \\
d_H\left[e^{4A-\phi}(\phmc)\right]+4i e^{3A-\phi}\I(\phpd) &= -\frac{1}{8}(\bar{\tilde{\xi}}\wedge+\iota_{\xi}) e^{4A} \ast\lambda(F) \label{F4} \ . 
\end{align}
\end{subequations}
The reason we have separated the last three equations is that they are in fact redundant given the ones above,\footnote{With the exception of the 0-form component of \eqref{F4} which does not involve the R--R fields due to $\xi$ being real.} as we will see in section \ref{RR}.
We also have
\begin{subequations}
\begin{align}
\I (\xi) &= 0 \ , \label{X1} \\
d(e^A f) + \frac{1}{2} \I(\tilde \xi)  &= 0 \ , \label{X2} \\
d \tilde \xi - i_\xi H &= 0 \ , \label{X3} \\
\nabla_{(n} \xi_{m)} &= 0 \label{X4} \ ,
\end{align}
\end{subequations}
which were obtained from \eqref{10d-eq2} and the condition that the ten-dimensional vector $K$ is Killing.

\subsection{Scalar and 1-form equations}
The 0-form components of \eqref{SUSYeq} and \eqref{F4} give:
\begin{subequations}\label{scalar_eq}
\begin{align}
a c  z_3^* &=-a \ , \\
(e^A+f) z_2^* &= (e^A-f) c z_1 \ ,
\end{align}
\end{subequations}
where in \eqref{F4} we have used \eqref{X1} to ``decouple'' $F$. 

Moving on to the 1-form equations, imposing \eqref{X1} yields
\begin{subequations}
\begin{align}
(1-|z_1|^2)V+1&=0 \ , \\
|a|^2c-V(z_3^*-z_1z_2)&=0 \ , \\
(1-|a|^2)cz_1^*-V(z_2-z_1^*z_3^*)&=0 \ ,
\end{align}
\end{subequations}
where
\begin{equation}
V\equiv\frac{a b c e^{i\vartheta}}{\sqrt{\det(Z)}} \ . 
\end{equation}
Combining the above with \eqref{scalar_eq} we arrive at
\begin{equation}
z_1=z_2=0 \ , \qquad z_3=-\frac{1}{c}=-|a| \ , \qquad V=-1 \ .
\end{equation}
The 1-form components of \eqref{SUSYeq} are then satisfied trivially and we are left with \eqref{X2}.

Henceforth, we will use the following parametrization for the scalars:
\begin{equation}
a = \cos\beta e^{i\alpha}, \qquad f = e^A \cos(2\theta) \ ,
\end{equation}
following the relations \eqref{abc} and \eqref{f_ineq}. 

\eqref{X2} now reads
\begin{equation}\label{dv2_1}
d\left(e^{2A}\cos(2\theta)\right) = - e^A \sin(2\theta) \R(w_1) \ .
\end{equation}

\subsection{Orthonormal frame and 1-form basis}
Before proceeding, we will introduce an orthonormal frame constructed out of $\{ w_1, w_2, w_3 \}$ and a new (non-orthogonal) 1-form basis that will prove useful in analysing the remaining supersymmetry equations.
The orthonormal frame is:
\begin{align}\label{ortho_frame}
\e^1 &= \frac{1}{\sin\beta} \left(\I(w_1) + \cos\beta \I(w_2) \right) \ , \qquad
\e^2 = \R(w_1) \ , \nn \\
\e^3 &= \I(w_2) \ , \qquad
\e^4 = \frac{1}{\sin\beta} \left(\R(w_2) + \cos\beta \R(w_1) \right) \ , \nn \\
\e^5 &= \R(w_3) \ , \qquad
\e^6 = \I(w_3) \ .
\end{align}
In terms of these, the SU(3) structure in (\ref{eq:JO}) reads
\begin{subequations}
\begin{align}
	J&= \e^1 \wedge (-\sin\beta \e^2 - \cos\beta \e^4)  + \e^3 \wedge (\cos\beta \e^2 - \sin\beta \e^4) 
    + \e^5 \wedge \e^6 \ ,\\
	\Omega &= e^{-i \alpha} 
	\left[ \e^1 + i (-\sin\beta \e^2 - \cos\beta \e^4) \right] \wedge 
	\left[ \e^3 + i (\cos\beta \e^2 - \sin\beta \e^4) \right] \wedge (\e^5 + i \e^6) \ .
\end{align}
\end{subequations}

This structure is the same that appeared in several IIA solutions: see for example \cite[Eq.~(3.6)]{petrini-zaffaroni}. There, it was identified from existing solutions (obtained by reduction  from eleven dimensions) and later imposed as an Ansatz. In our approach, it is coming out naturally.

The new basis is:
\begin{subequations}\label{basis}
\begin{align}
v^1 &= 2 (e^A \sin \beta \sin(2 \theta))^{-1} \, \e^1 \ , \\ 
v^2 &= -e^A \sin(2\theta) \, \e^2 \ , \\ 
v^3 &= e^{3A-\phi}\left(\sin\alpha\cos\beta\cos(2\theta) \, \e^2-\cos\alpha \, \e^3 + \sin\alpha\sin\beta \, \e^4\right) 
\ , \\
v^4 &= (e^{A+\phi} \cos(2\theta))^{-1}\left(\cos\alpha\cos\beta(\cos(2\theta))^{-1} \, \e^2 + \sin\alpha \, \e^3 + \cos\alpha\sin\beta \, \e^4 \right) \ , \\
v^5 &=-2e^{3A-\phi}\sin\beta\sin(2\theta) \, \e^5 \ , \\
v^6 &=-2e^{3A-\phi}\sin\beta\sin(2\theta) \, \e^6 \ .
\end{align}
\end{subequations}
We will also use
\begin{equation}
v = (e^A\sin(2\theta))^{-1}\left(\cos(2\theta) \, \e^2 - \cot\beta \, \e^4\right) \ ,
\end{equation}
or expressed in terms of the ``$v$ basis'' (\ref{basis}):
\begin{align}\label{v}
v &= - \left(\frac{\cos(2\theta)}{e^{2A}\sin^2(2\theta)} + \frac{\cos^2\beta\sin^2\alpha\cos^2(2\theta)+\cos^2\beta\cos^2\alpha}{e^{2A}\sin^2\beta\sin^2(2\theta)\cos(2\theta)}\right) v^2 \nn \\
& \hspace{2cm} - \frac{e^{-4A+\phi}\sin\alpha\cos\beta}{\sin^2\beta\sin(2\theta)} v^3 
 - \frac{e^\phi\cos\beta\cos\alpha\cos(2\theta)}{\sin^2\beta\sin(2\theta)} v^4 \ .
\end{align}

\subsection{2-form equations}
From the 2-form components of \eqref{SUSYeq} we get:
\begin{equation}\label{dv56}
d(v^5+iv^6) = (2 v -i v^1) \wedge (v^5+iv^6) \ , 
\end{equation}
and 
\begin{equation}\label{dv3dv4}
dv^3 = 0 \ , \qquad dv^4 = 0 \ .
\end{equation}

We are left with \eqref{X3} which we will rewrite as
\begin{equation}\label{H}
 H = - \frac{1}{4} d\left(v^1 \wedge \R(\tilde{\xi})\right) + H_0 \ ,
\end{equation}
using the fact that $\xi = 2 e^A \sin \beta \sin(2 \theta) \, (\e^1)^\sharp$, where $(\e^1)^\sharp$ denotes the vector dual to $\e^1$.

\subsection{$(p > 2)$-form equations}
The 3-form components of $\eqref{SUSYeq}$ are satisfied trivially given the results derived so far, whereas the 4-form components yield the conditions
\begin{equation}\label{hf1}
\left(d \delta_1 + i H_0 \right)
\wedge (v^5 + i v^6) = 0 \ ,
\end{equation}
and
\begin{align}
\left( d \delta_2 + 4 v \wedge \delta_2 \right) \wedge v^{56}  
&= 2 e^{4A} \cos^2(2\theta) v^{24} \wedge dv^1 + 4H_0 \wedge v^3  \label{hf2} \ , \\
\left( d \delta_3 + 4 v \wedge \delta_3 \right) \wedge v^{56}  
&= 2 v^{23} \wedge dv^1 - 4 e^{4A} \cos^2(2\theta)H_0 \wedge v^4 + \frac{2e^{-8A+2\phi}}{\sin^2\beta\sin^2(2\theta)} v^{23} \wedge v^{56} \label{hf3} \ .
\end{align}
In the above
\begin{subequations}\label{deltas}
\begin{align}
\delta_1 &\equiv \frac{1}{\sin\beta} \e^3 \wedge \e^4 \ , \\
\delta_2 &\equiv \frac{e^{-3A+\phi}}{\sin^2\beta\sin^2(2\theta)}\left(e^{A+\phi}\cos(2\theta) v^4 
+ e^{-A}\cos\alpha\cos\beta\tan(2\theta) v^2
\right) \ , \\
\delta_3 &\equiv \frac{e^{-3A+\phi}}{\sin^2\beta\sin^2(2\theta)}\left(e^{-3A+\phi}\cos(2\theta) v^3 
- e^{-A}\sin\alpha\cos\beta\sin(2\theta)  v^2
\right) \ .
\end{align}
\end{subequations}
Also $v^{56} \equiv v^5 \wedge v^6$ etc.

Finally, the 5-form and 6-form components of $\eqref{SUSYeq}$, given the conditions derived so far, are trivially satisfied.

\subsection{Equations with R--R fields}
\label{RR}
Out of the equations which involve the R--R fields, only \eqref{F2} is independent, with \eqref{F1}, \eqref{F3} and \eqref{F4} following from it given the rest of the supersymmetry equations. 

Here is a sketch, for example, of how to show that (\ref{F3}) is redundant. One can act with $\bar{\tilde{\xi}} \wedge + \iota_\xi$  on (\ref{F2}). The right-hand side of (\ref{F2}) now becomes proportional to the right-hand side of (\ref{F3}). For the left-hand side we can use  
\begin{equation}
	\{\bar{\tilde{\xi}} \wedge + \iota_\xi , d_H\} = (d\bar{\tilde{\xi}} - \iota_\xi H )\wedge + L_\xi  = L_\xi \ ,
\end{equation}
the Lie derivative under $\xi$, where the last equality follows from (\ref{X3}). The action of $L_\xi$ on the pure spinors is the one dictated by their total R-charge: $L_\xi \phi^{\pm \pm}_+=0$, $L_\xi \phi^{\pm \mp}_-=0$, and $\phi^{\pm \mp}_+$, $\phi^{\pm\pm}_-$ have charges $\pm 2$.
 Using several Fierz identities one can show 
\begin{equation}
\begin{split}
	&(\bar{\tilde{\xi}} \wedge + \iota_\xi) (\phi^{++}_+ - \phi^{--}_+)=  (\bar{\tilde{\xi}} \wedge + \iota_\xi) (\overline{\phi^{++}_+} - \overline{\phi^{--}_+})= -4i (e^A \phi^{(+-)} + f \phi^{[+-]}) \ , \\
	&(\bar{\tilde{\xi}} \wedge + \iota_\xi) \phi^{[+-]}_- = 0 \ , \qquad
	(\bar{\tilde{\xi}} \wedge + \iota_\xi) \overline{\phi^{[+-]}_-} = -2i e^A{\rm Re} (\phi^{++}_+ + \phi^{--}_+) - 2 f {\rm Im} (\phi^{++}_+-\phi^{--}_+)\ . 
\end{split}	 
\end{equation}
Using also (\ref{X2}), one can now massage the result to obtain (\ref{F3}). A similar argument shows that (\ref{F4}) follows from (\ref{F1}). 

In spite of being redundant, \eqref{F1} and \eqref{F3} are useful for showing in a straightforward way that the equations of motion and the Bianchi identities of the R--R fields are automatically satisfied.

Acting with $d_H$ on \eqref{F1}, and using the imaginary part of \eqref{S2} it follows that
\begin{equation}
d_H(e^{4A}*\lambda(F)) = 0 \ , 
\end{equation}
which are the equations of motion. 

Acting with $d_H$ on \eqref{F2}, using \eqref{X2}, and subtracting the real part of \eqref{F3}, it follows that
\begin{equation}
d_H F = 0 \ , 
\end{equation}
which are the Bianchi identities of the R--R fields.

Finally, equation \eqref{F2} determines the R--R fields. We give their expressions in section \ref{total}.

\subsection{Summary}
\label{sub:sum1}

We have formulated the supersymmetry equations as a set of differential constraints on an identity strucure parametrized by the set of functions $\{A, \phi, \theta, \alpha, \beta\}$ and the 1-forms \eqref{basis}, which are subject to
\begin{align}
v^2 = d\left(e^{2A}\cos(2\theta)\right) \ , \qquad dv^3 = 0 \ , \qquad dv^4 = 0 \ , \label{dv2dv3dv4} \\
d(v^5+iv^6) = (2 v -i v^1) \wedge (v^5+iv^6) \label{dv5v6} \ , 
\end{align}
(with $v$ given by \eqref{v}), as well as \eqref{hf1}, \eqref{hf2} and \eqref{hf3}. Finally, 
$\xi = \frac{1}{4}||\xi||^2 (v^1)^\sharp$ (where ${}^\sharp$ denotes raising the index) is a Killing vector. 
In the next section we will refine the analysis of these constraints by introducing coordinates, thus reducing them to partial differential equations.

The NS--NS field strength is given by \eqref{H}, with $H_0$ determined by \eqref{hf1}--\eqref{hf2}; we will give its explicit expression in the next section. The R--R field strengths are given by \eqref{F2}. Note that the Bianchi identities for the form fields need to be imposed on top of the supersymmetry equations. However, as we saw in section 
\ref{RR}, the Bianchi identities for the R--R fields are already implied by the latter. The Bianchi identity for $H$ still
needs to be imposed and we will do so in the next section.

\section{Local coordinates and partial differential equations}
\label{sec:local}

In this section we introduce local coordinates and a new set of functions that will allow us to solve some of the
conditions derived in the previous section, and reduce the rest to a system of partial differential equations.

\subsection{Local coordinates and a new set of functions}

We start by introducing the coordinates $\{ y \equiv e^{2A} \cos(2\theta), \lambda_1, \lambda_2 \}$ so that \eqref{dv2dv3dv4} are solved as
\begin{equation}
v^2 = dy \ , \qquad v^3 = d\lambda_1 \ , \qquad v^4 = d\lambda_2 \ .
\end{equation} 
Next, we introduce the coordinate $\psi$ adapted to the Killing vector $\xi$:
\begin{equation}
\xi = 4 \partial_\psi \ .
\end{equation}
It follows that
\begin{equation}\label{v1}
v^1 = d\psi + \rho \ ,
\end{equation}
for a 1-form $\rho$. 

Finally, the differential equation \eqref{dv5v6} can be solved by
\begin{equation}\label{eq:v56dx12}
v^5 + i v^6 = e^{-i\psi} e^{2\Sigma}(dx_1 + i dx_2) \ ,
\end{equation}
for a function $\Sigma = \Sigma(y, \lambda_1, \lambda_2, x_1, x_2)$. 
We give a detailed explanation of this in appendix \ref{v5v6coord}, but a summary is that one needs the ``complex Frobenius theorem'' by Nirenberg \cite{nirenberg}, which is a mix between the real Frobenius theorem and the Newlander--Nirenberg theorem about integrability of complex structures. In general it says the following: let $M$ be a manifold of dimension $n$. Given a subbundle $\Omega \subset (T^*M)^{\Bbb C}$ of dimension $k$, and $\Lambda\equiv \Omega \cap \bar \Omega$ of dimension $k'$, then there exist locally adapted coordinates such that $\Omega$ is spanned by $dx_a+ i dx_{a+l}$, $a=1,\ldots,l\equiv k-k'$ and $dx_\sigma$, $\sigma=n-k'+1,\ldots,n$, if and only if $d \Omega\subset $ the ideal generated by $\Omega$, and $d \Lambda \subset$ the ideal generated by $\Lambda$. It is used in the theory of transversely holomorphic foliations (THF); see for example \cite[Thm.~1.8]{duchamp-kalka}.\footnote{In physics, a THF appears for example as a condition on which three-manifolds preserve at least one supercharge of a supersymmetric field theory \cite{ktz,Closset:2012ru}, with the only difference that the leaves there are one-dimensional. Another physics application is to A-branes \cite{kapustin-orlov}. Finally, the logic explained here was also used (implicitly) in \cite{bah-passias-t}.} In our case, we can take $\Omega$ to be the span of $v^5+iv^6$; $\Lambda=\{0\}$. Then the condition  $d \Omega\subset $ the ideal generated by $\Omega$ is simply (\ref{eq:dw3aw3}). This implies that there are adapted coordinates such that $\Omega$ is the span of $dx_1+idx_2$.

With (\ref{eq:v56dx12}), \eqref{dv5v6} now yields
\begin{equation}\label{diff_v}
v = \Sigma_{,y} dy + \Sigma_{,\lambda_1} d\lambda_1 + \Sigma_{,\lambda_2} d\lambda_2 \ , 
\end{equation}
and
\begin{equation}\label{rho}
\rho = - 2 \Sigma_{,x_2} dx^1 + 2 \Sigma_{,x_1} dx^2 \ .
\end{equation}
Here $\Sigma_{,y} \equiv \partial_y \Sigma$ etc..

Via the two expressions for $v$, \eqref{v} and \eqref{diff_v}, we can exhange some of the functions we have been using in the supersymmetry equations with derivatives of $\Sigma$. 
In particular
\begin{subequations}
\begin{align}
\Sigma_{,\lambda_1} &= - \frac{e^{-4A+\phi}\sin\alpha\cos\beta}{\sin^2\beta\sin(2\theta)} \ , \\
\Sigma_{,\lambda_2} &= - \frac{e^\phi\cos\beta\cos\alpha\cos(2\theta)}{\sin^2\beta\sin(2\theta)} \ ,  \\
\Sigma_{,y} &= -\frac{\cos(2\theta)}{e^{2A}\sin^2(2\theta)} - \frac{\cos^2\beta\sin^2\alpha\cos^2(2\theta)+\cos^2\beta\cos^2\alpha}{e^{2A}\sin^2\beta\sin^2(2\theta)\cos(2\theta)}
\label{sigma_y} \ .
\end{align}
\end{subequations}
By also introducing
\begin{equation}
\Lambda = \frac{e^{-2A+2\phi}\cos(2\theta)}{\sin^2\beta} \ , 
\end{equation}
we can express $\{A, \phi, \theta, \alpha, \beta\}$ in terms of 
$\{y, \Sigma_{,y}, \Sigma_{,\lambda_1}, \Sigma_{,\lambda_2}, \Lambda\}$, of which $y$ is used as a coordinate, thus reducing the number of functions that characterize the solutions to two: $\Sigma$ and $\Lambda$.
Explicitly,
\begin{align}\label{func}
e^{4A} &= \frac{\Lambda}{U} + y^2 \ , \qquad 
e^{2\phi} = - \frac{(\Lambda + y^2 U)^2}{(y^{-1}\Sigma_{,\lambda_2})^2 + \Sigma_{,y} (\Lambda + y^2 U)} \ , \nn \\
\cos(2\theta) &= y \left(\frac{\Lambda}{U} + y^2\right)^{-1/2} \ , \qquad
\tan\alpha = \frac{y\Sigma_{,\lambda_1}}{\Sigma_{,\lambda_2}} \left(\frac{\Lambda}{U} + y^2\right)^{1/2} \ , 
 \nn \\
\cot^2(\beta) &= \frac{(\Sigma_{,\lambda_2})^2}{y(\Lambda + y^2 U)} + \frac{y}{U}(\Sigma_{,\lambda_1})^2  \ . 
\end{align}
where
\begin{equation}\label{U}
U \equiv - y^{-1}(\Sigma_{,y}\Lambda +(y\Sigma_{,\lambda_1})^2 + (y^{-1}\Sigma_{,\lambda_2})^2)
\end{equation}
is not an independent function, but will be convenient to use.

In the following section we will reduce the rest of the supersymmetry conditions to a set of partial differential equations for $\Sigma$ and $\Lambda$.

\subsection{Partial differential equations}
Before moving on with the analysis of the supersymmetry equations, we define the Hodge star operators $*_x$:
\begin{equation}
*_x dx_1 = dx_2, \qquad *_x dx_2 = - dx_1 \ , 
\end{equation}
and $*_\lambda$:
\begin{equation}
*_\lambda d\lambda_1 = y^2 d\lambda_2, \qquad *_\lambda d\lambda_2 = - y^{-2} d\lambda_1 \ , 
\end{equation}
and the corresponding Laplacians
\begin{equation}
\Delta_x = \partial^2_{x_1} + \partial^2_{x_2} \ , \qquad
\Delta_\lambda = y^2 \partial^2_{\lambda_1} + y^{-2} \partial^2_{\lambda_2} \ .
\end{equation}
We will also use $d_\lambda \equiv d\lambda_1 \wedge \partial_{\lambda_1} + d\lambda_2 \wedge \partial_{\lambda_2}$ and
$d_x \equiv dx_1 \wedge \partial_{x_1} + dx_2 \wedge \partial_{x_2}$.

The supersymmetry conditions to analyze are \eqref{hf1}, \eqref{hf2} and \eqref{hf3}; they will yield two partial differential equations for $\{\Sigma, \Lambda\}$ and an expression for $H_0$. In terms of the coordinates and the new functions:
\begin{subequations}
\begin{align}
\delta_1 &= dy \wedge *_{\lambda} d_\lambda \Sigma 
- \Lambda \, d\lambda_1 \wedge d\lambda_2 \ , \label{delta1} \\
\delta_2 &= \left(y^2 U + \Lambda \right) d\lambda_2 - y^{-2} \Sigma_{,\lambda_2} dy \ , \\
\delta_3 &= U d\lambda_1 + \Sigma_{,\lambda_1} dy \ .
\end{align}
\end{subequations}

Let us start with the differential equations. \eqref{hf1} gives
\begin{equation}\label{MEQ1}
\Delta_\lambda \Sigma = - \Lambda_{,y} , 
\end{equation}
which combined with \eqref{U} can be alternatively written as 
\begin{align}\label{MEQ1alter}
\Delta_\lambda e^{4\Sigma} = - 4 (e^{4\Sigma} \Lambda)_{,y} - 16 e^{4\Sigma} y U \ .
\end{align}
\eqref{hf2} and \eqref{hf3} give two expressions for the $(x_1,x_2)$ components of $dv^1$
\begin{subequations}
\begin{align}
(dv^1)|_{x_1x_2} &= \frac{1}{2} \left( e^{4\Sigma} U \right)_{,y} + \frac{1}{y} e^{4\Sigma} U
+ \frac{1}{2y^2} \left[ \frac{1}{4} y^{-2} (e^{4\Sigma})_{,\lambda_2\lambda_2} + (e^{4\Sigma} \Lambda)_{,y} \right] \ , \\
(dv^1)|_{x_1x_2} &= \frac{1}{2} \left( e^{4\Sigma} U \right)_{,y} 
 - \frac{1}{y} e^{4\Sigma} U
 - \frac{1}{8} (e^{4\Sigma})_{,\lambda_1\lambda_1} \ , 
\end{align}
\end{subequations} 
which given \eqref{MEQ1alter} can be shown to be equivalent. Combining these with \eqref{v1} and \eqref{rho} we obtain the equation
\begin{equation}\label{MEQ2}
\Delta_x \Sigma  + \frac{1}{16} (e^{4\Sigma})_{,\lambda_1\lambda_1} = \frac{1}{4} y^2 \left( e^{4\Sigma} y^{-2} U \right)_{,y} \ .
\end{equation}

Turning to $H_0$, \eqref{hf1} determines its 
$\{(y,\lambda_{1,2},x_{1,2}),(\lambda_1,\lambda_2,x_{1,2})\}$ components, while \eqref{hf2} and \eqref{hf3} its
$\{(y,x_1,x_2),(\lambda_{1,2},x_1,x_2)\}$ components. In total we get:
\begin{align}\label{H0}
H_0 &= \frac{1}{2} dy \wedge *_\lambda d_\lambda  \rho 
- d\lambda_1 \wedge d\lambda_2 \wedge *_x d_x \Lambda \\ \nn 
&+ \left[ \frac{1}{16} y^{-2} (e^{4\Sigma})_{,\lambda_2\lambda_1} dy  - \frac{1}{4} *_\lambda d_\lambda (e^{4\Sigma} U) - \frac{1}{4} \left(e^{4\Sigma}\Lambda\right)_{,\lambda_1} d\lambda_2 \right] \wedge dx_1 \wedge dx_2
\end{align}
Having fully specified $H$, via \eqref{H} and \eqref{H0} we can impose its Bianchi identity, $dH=0$. By doing so we get
\begin{equation}\label{MEQ3}
\Delta_x \Lambda = - \frac{1}{4} \Delta_\lambda (e^{4\Sigma}U) - \frac{1}{4} (e^{4\Sigma}\Lambda)_{,\lambda_1\lambda_1} \ .
\end{equation}

\section{Summary of final results}\label{total}
We have reduced the proplem of finding supersymmetric AdS$_4$ solutions to solving three partial differential equations (PDEs) for two functions $\Sigma$ and $\Lambda$ of five variables $\{y, \lambda_1, \lambda_2, x_1, x_2\}$:
\begin{subequations}\label{eq:total}
\begin{empheq}[box=\fbox]{align}
\Delta_\lambda \Sigma = - \Lambda_{,y} 
\, , \label{PDE-1} \\
\Delta_x \Sigma  + \frac{1}{16} (e^{4\Sigma})_{,\lambda_1\lambda_1} = \frac{1}{4} y^2 \left( e^{4\Sigma} y^{-2} U \right)_{,y} 
\, , \label{PDE-2} \\
\Delta_x \Lambda = - \frac{1}{4} \Delta_\lambda (e^{4\Sigma}U) - \frac{1}{4} (e^{4\Sigma}\Lambda)_{,\lambda_1\lambda_1} 
\, , \label{PDE-3} \\[.2cm]
U \equiv - y^{-1}(\Sigma_{,y}\Lambda +(y\Sigma_{,\lambda_1})^2 + (y^{-1}\Sigma_{,\lambda_2})^2)  \, , 
\end{empheq}
\end{subequations}
where 
\begin{equation}\label{eq:Lap}
	\Delta_x = \partial^2_{x_1} + \partial^2_{x_2} \, ,\qquad
	\Delta_\lambda = y^2 \partial^2_{\lambda_1} + y^{-2} \partial^2_{\lambda_2}\,.
\end{equation}

By inverting \eqref{basis} so that the orhonormal frame is expressed in terms of the $v$'s, and eventually in terms of the
coordinates introduced in the previous section, we can write down the metric for $M_6$:
\begin{equation}
\begin{split}
ds^2_6 &= e^{-6A+2\phi} U^{-1} 
\left\{ 
y \left[\frac{1}{4}(d\psi+\rho)^2 + (v)^2\right] +
U d\lambda_1^2
+  y^2 e^{4A} U d\lambda_2^2
-2\Sigma_{,\lambda_2} \, dy \, d\lambda_2 - \Sigma_{,y} \, dy^2
\right\} \\
&+ \frac{1}{4} y^{-1}U e^{4\Sigma+2A}(dx_1^2 + dx_2^2) \ ,
\end{split}
\end{equation}
where the warp function $A$ and the dilaton $\phi$ are given by 
\begin{align}
e^{4A} &= \frac{\Lambda}{U} + y^2 \ , \qquad 
e^{2\phi} = - \frac{(\Lambda + y^2 U)^2}{(y^{-1}\Sigma_{,\lambda_2})^2 + \Sigma_{,y} (\Lambda + y^2 U)} \ ,
\end{align}
while
$v = \Sigma_{,y} dy + \Sigma_{,\lambda_1} d\lambda_1 + \Sigma_{,\lambda_2} d\lambda_2$ and
$\rho = - 2 \Sigma_{,x_2} dx^1 + 2 \Sigma_{,x_1} dx^2$.
$M_6$ has a transversely holomorphic foliation of codimension 1, with the coordinates on the leaves being $\{\psi, y, \lambda_1, \lambda_2\}$. There is a U$(1)$ isometry acting on $\psi$, which is a symmetry of the full solution, and corresponds to the R-symmetry of the dual superconformal field theory. Moreover, the $\psi$ circle is fibered over the surface parameterized by $\{x_1,x_2\}$.

The NS--NS field reads
\begin{align}
H = &- \frac{1}{4} d\left((d\psi+\rho) \wedge \R(\tilde{\xi})\right) + \frac{1}{2} dy \wedge *_\lambda d_\lambda  \rho 
- d\lambda_1 \wedge d\lambda_2 \wedge *_x d_x \Lambda \\ \nn 
&+ \left[ \frac{1}{16} y^{-2} (e^{4\Sigma})_{,\lambda_2\lambda_1} dy  - \frac{1}{4} *_\lambda d_\lambda (e^{4\Sigma} U) - \frac{1}{4} \left(e^{4\Sigma}\Lambda\right)_{,\lambda_1} d\lambda_2 \right] \wedge dx_1 \wedge dx_2 \ ,
\end{align}
where
\begin{equation}
\R(\tilde{\xi}) = -2 \frac{e^{-8A+2\phi}}{U^2}
\left(\Sigma_{,\lambda_1}\Sigma_{,\lambda_2} dy
- y^2 e^{4A} U \Sigma_{,\lambda_1} d\lambda_2 + U \Sigma_{,\lambda_2} d\lambda_1 
 \right) \ .
\end{equation}

The R--R fields read:
\begin{subequations}
\begin{align}
F_1 &= df_0 + d\lambda_2 \ , \\
F_3 &= d_+ f_2 - H_+ f_0  + f_3 \ , \\
F_5 &= d_+ f_4 - H_+ \wedge f_2 \ , 
\end{align}
\end{subequations}
where
\begin{equation}
d_+ \equiv d + \frac{e^{4A}U}{y\Lambda} dy \wedge \ , \qquad 
H_+ \equiv H + \frac{1}{2} \frac{e^{-4A+2\phi}y}{\Lambda} (d\psi+\rho) \wedge \delta_1 \ ,
\end{equation}
with $\delta_1$ given by \eqref{delta1}, and
\begin{subequations}
\begin{align}
f_0 &\equiv  \frac{\Sigma_{,\lambda_2}}{ye^{4A}U} \ , \\
f_2 &\equiv -\frac{1}{2} \frac{e^{-8A+2\phi}}{U^2} (d\psi+\rho) \wedge 
\left(U d\lambda_1 + y\Sigma_{,\lambda_1}v\right) 
-\frac{1}{4y} \Sigma_{,\lambda_1} \vol_x \ , \\
f_3 &\equiv \frac{1}{4} U (\Lambda^{-1}\Sigma_{,\lambda_1}dy-y^{-2}d\lambda_1) \wedge \vol_x \ , \\
f_4 &\equiv \frac{1}{8} \frac{e^{-4A}}{\Sigma_{,\lambda_2}} 
\left[e^{4A}U dy  - e^{2\phi} \Sigma_{,\lambda_1} d\lambda_1
+ e^{2\phi} \left(1 + \frac{(\Sigma_{,\lambda_2})^2}{ye^{4A}U}\right)
v \right]
\wedge d\psi \wedge \vol_x \ ,
\end{align}
\end{subequations}
where $\vol_x = e^{4\Sigma} dx_1 \wedge dx_2$.

\section{Solutions}
\label{sec:sol}

\subsection{AdS$_5 \times S^5$}
In this section we recover the AdS$_5 \times S^5$ solution from our system of equations, by imposing that the $(\psi, x_1, x_2)$ subspace forms a round three-sphere, as well as constant axion ($F_1 = 0$) and dilaton. It will be convenient to work with the functions $\{\Sigma,A\}$.
The first condition amounts to 
\begin{equation}
\Sigma = \frac{1}{2} A_0(x_1,x_2) + s(y,\lambda_1,\lambda_2) \ , \qquad \Delta_x A_0 = - e^{2A_0}
\end{equation}
and
\begin{equation}\label{con-1}
e^{4A} U = y g_s e^{-2s} \ , \qquad g_s = e^{\phi} = {\rm const.} \ .
\end{equation}
Requiring that $F_1 = 0$ and constant dilaton gives respectively:
\begin{subequations}
\begin{align}
e^{4A} U = \frac{s_{,\lambda_2}}{y(C_0-\lambda_2)} \ , \label{con-2} \\
g_s^2 = - \frac{e^{8A}U^2}{(y^{-1}s_{,\lambda_2})^2 + e^{4 A}U s_{,y}}
\ , \label{con-3}
\end{align}
\end{subequations}
where $C_0$ is constant. In what follows, by shifting $\lambda_2$, we will set it to zero. 

Combining \eqref{con-1}, \eqref{con-2}, and \eqref{con-3} together with $e^{4A}(U,\Lambda,y)$ from \eqref{func} and $U(\Sigma,\Lambda,y)$ from \eqref{U} we arrive at 
\begin{equation}
e^{2s} = -y^2(g_s^{-1}+g_s\lambda_2^2) + h(\lambda_1) \ , 
\end{equation}
with $h(\lambda_1)$ satisfying 
\begin{equation}\label{ode-1}
\left(\frac{dh}{d\lambda_1}\right)^2 = 4(1 - g_s e^{-4A} h) \ .
\end{equation}
From the latter equation we conclude that $A=A(\lambda_1)$.

Finally we need to solve the PDEs that comprise our system of equations.  Starting with \eqref{PDE-3}, 
we find that it gives
\begin{equation}
\frac{d^2h}{d\lambda_1^2} = 2 g_s e^{-4A} \ .
\end{equation}
Combining the above with \eqref{ode-1} we get
\begin{align}
h &= c_0 e^{2A}\ , \qquad c_0 = {\rm const.} \ , \\
d\lambda_1 &= \pm \frac{c_0 e^{2A}}{\sqrt{1-g_s c_0 e^{-2A}}} dA \ .
\end{align}
The rest of the PDEs, \eqref{PDE-1} and \eqref{PDE-2}, are then automatically satisfied. 

Turning to the internal metric we write it as
\begin{equation}
g_s^{-1} ds^2_{6} = e^{2s-2A}\left(ds^2_{S^3} + ds^2\right) + 
\frac{g_s c^2_0 e^{-2A}}{1-g_s c_0 e^{-2A}} dA^2
+ \frac{c_0-e^{2s-2A}}{(g_s^{-1}+g_s\lambda_2^2)^2} d\lambda_2^2 + e^{-2A} (d\sqrt{c_0 e^{2A}-e^{2s}})^2 , 
\end{equation}
effectively switching coordinates from $\{y,\lambda_1\}$ to $\{s,A\}$. Here
\begin{equation}
ds^2_{S^3} = \frac{1}{4}\left[(d\psi+\rho)^2 + e^{2A_0}(dx_1^2 + dx_2^2)\right] \ , 
\end{equation}
is the metric on the round three-sphere, of unit radius.
Introducing new coordinates $\{x,\phi_1,\phi_2\}$ via
\begin{equation}
A = \log(\sqrt{g_s c_0} \cosh \varrho) \ , \qquad
e^{2s-2A} = c_0 \sin^2(\phi_1) \ , \qquad 
\phi_2 = \arctan(g_s \lambda_2) \ ,
\end{equation}
the ten-dimensional metric becomes the AdS$_5 \times S^5$ metric
\begin{equation}
ds^2_{10} = L^2 \left( d\varrho^2 + \cosh^2(\varrho) ds^2_{\rm AdS_4} + d\phi_1^2 + \sin^2(\phi_1) ds^2_{S^3} + \cos^2(\phi_1) d\phi_2^2 \right) \ ,
\end{equation}
with $L^2 = g_s c_0$ and $ds^2_{\rm AdS_5}=d\varrho^2 + \cosh^2(\varrho) ds^2_{\rm AdS_4}$.

Finally, looking at the form fields, as expected $F_3$ and $H$ are zero, whereas
\begin{equation}
F_5 = 4 g_s c_0^2 \vol_{S^5} \ . 
\end{equation}
Flux quantization
\begin{equation}
N \equiv \frac{1}{16\pi^4\alpha'^2} \int F_5 = \frac{g_s c_0^2}{4\pi\alpha'^2}
\end{equation}
gives
\begin{equation}
c_0^2 = 4\pi \alpha'^2 N g_s^{-1} \ , 
\end{equation}
and hence, 
\begin{equation}
L^2  =  \alpha' \sqrt{4 \pi g_s N} \ , \qquad F_5 = 16 \pi \alpha'^2 N \vol_{S^5} \ .
\end{equation}

\subsection{Separation of variables Ansatz} 
\label{sub:sep}

We will now discuss an Ansatz that allows for several classes of new solutions. It involves the natural assumption that the two-dimensional surface parameterized by $\{x_1, x_2\}$ is a Riemann surface of constant curvature. As we warned in the introduction, we have not pursued a global analysis to the point of making sure there is a class for which the internal space is compact and physical. However, new solutions seem to be generated easily enough that this is likely to be achieved. We expect to report on this in the future.

The Ansatz consists of 
\begin{equation}\label{eq:sep}
	\Sigma= \frac{1}{2} A_0 (x_1,x_2) + s(y, \lambda_1, \lambda_2) \ ,\qquad
	\Lambda = \Lambda(y, \lambda_1, \lambda_2) \ ,
\end{equation}
where $A_0$ is a solution of Liouville's equation
\begin{equation}
	\Delta_x A_0 = - \kappa e^{2A_0} \ ,\qquad \kappa \in \{-1,0,1\} \ ,
\end{equation}
and can be taken to be $A_0=- \log((1+\kappa(x_1^2 + x_2^2))/2)$.

It proves useful to define
\begin{equation}
	 E\equiv e^{4s} \ ,\qquad V \equiv U e^{4s} \ ,\qquad
	L \equiv \Lambda e^{4s} \ .
\end{equation}
With these definitions the system (\ref{eq:total}) becomes\footnote{The first equation is a modification of the corresponding one in (\ref{eq:total}) using the rest.}
\begin{equation}\label{eq:sep-total}
\begin{split}
	&(L+y^2 V)_{,y}= -\frac1{4} y^{-2} E_{,\lambda_2 \lambda_2} - 2 \kappa y^2
	\ ,\qquad
	-2\kappa= y^2 (y^{-2}V)_{,y}-\frac14 E_{,\lambda_1 \lambda_1} \ , \\
	&(L+ y^2 V)_{,\lambda_1 \lambda_1} = -y^{-2}V_{,\lambda_2 \lambda_2} \ ,\qquad
	-yVE =  \frac14 L E_{,y} + \frac{1}{16}\left(y E_{,\lambda_1} \right)^2 + \frac{1}{16}\left(y^{-1} E_{,\lambda_2} \right)^2 \ .
\end{split}
\end{equation}
Notice that three of the above equations are linear in $E$, $V$ and $L$, and only one is quadratic. This feature makes it easier to find solutions.

The dilaton and metric become
\begin{align}
	&e^{-2\phi}= 
	-\frac{1}{e^{8A} V^2} \left( \frac1{16}(y^{-1} E_{,\lambda_2})^2 + \frac14 e^{4A} V E_{,y}\right) \, ; \\
	&ds^2_6= \frac{e^{2A}V}{4y} ds^2_{\mathcal{C}} + \frac{e^{-6A+2 \phi}}{4V} \left[y \left(E D \psi^2 + (d\sqrt{E})^2\right) + 4V\left(d \lambda_1^2 + y^2 e^{4A}d \lambda_2^2\right) - 2E_{,\lambda_2} d \lambda_2 d y - E_{,y} d y^2 \right] \, , 
	 \label{eq:sep-metric}
\end{align}	
where $e^{4A}= L/V + y^2$, $ds^2_\mathcal{C} = e^{2A_0}(dx_1^2 + dx_2^2)$ is the line element of a Riemann surface of scalar curvature $2\kappa$, and $D\psi \equiv d\psi + \rho$.
Here the coordinates are $\{ x_i, y, \lambda_i \}$, $i=1,2$, with $dE = E_{,y} dy + E_{,\lambda_i}  d \lambda_i $. 
As can be seen, at the locus where $E$ goes to zero, the $\psi$ circle shrinks regularly by fixing the period of $\psi$ to be $2\pi$.

One can also eliminate $y$ as a coordinate in favor of $E$. This leads to the alternative expression for the metric: 
\begin{align}\label{eq:sep-metric2}
ds^2_6 &= e^{2A}\frac V{4y} ds^2_{\mathcal{C}} + \frac{e^{-6A+2 \phi}}{4V} ds^2_{4} \ , \nn \\
ds^2_4 &= yED \psi^2 + \left(\frac{y}{4 E} - y_{,E}\right)dE^2 - 2y_{,\lambda_1} dE d \lambda_1 +\left(4V-\frac{(y_{,\lambda_1})^2}{y_{,E}}\right) d \lambda_1^2 + \left(4 V y^2 e^{4A}+\frac{(y_{,\lambda_2})^2}{y_{,E}}\right)  d\lambda_2^2 \ .
\end{align}
Although this expression appears longer, it has the advantage of having fewer non-diagonal components.

We will now explore two classes of sub-Ans\"atze. 

\subsubsection{Compactification Ansatz} 
\label{ssub:comp}

The first class comes about by demanding that the line element of the Riemann surface $\mathcal{C}$ has the same prefactor as that of AdS$_4$, so that the metric takes the form $ds^2_{10} = e^{2A}(ds^2_{{\rm AdS}_4}  + \tfrac{1}{4} ds^2_\mathcal{C}) + \ldots$ . The holographic interpretation of this class of solutions is that of the dual of a five-dimensional field theory compactified on $\mathcal{C}$. An analogous class was studied in \cite{afpt}, where the so-called ``compactification Ansatz'' was applied to AdS$_5$ solutions. 

From (\ref{eq:sep-metric}) we see that this Ansatz amounts to imposing 
\begin{equation}\label{eq:Vry}
	V= r y\,
\end{equation}
where $r$ is a constant proportional to the curvature radius of $\mathcal{C}$. 

With (\ref{eq:Vry}), two of the equations in (\ref{eq:sep-total}) determine 
\begin{equation}
	E= 2(2\kappa-r) \lambda_1^2 + K_1 \lambda_1 + K_2 \, ,\qquad 
	L=  L_1 \lambda_1 + L_2 \, ,
\end{equation}
where $K_i$ and $L_i$ are functions of $\lambda_2$ and $y$. In the remaining two equations, $\lambda_1$ only appears linearly or quadratically; thus one can expand in it, and obtain several PDEs in $\lambda_2$ and $y$ only. Some of them are quadratic, but further assumptions make them manageable. For example one may impose that $K_i$ and $L_i$ do not depend on $\lambda_2$. The most ``physically promising'' solution one finds like this is
\begin{equation}
	E= -8 \lambda_1^2 + k_1 \lambda_1 - \frac{1}{32} k_1^2 \, ,\qquad 
	L=  \ell_1 \lambda_1 + \ell_2-\frac43 y^3 \, ,\qquad r=-2\kappa=2\,,
\end{equation}
where $k_i$, $\ell_i$ are constant. More complicated solutions exist; for example: 
\begin{equation}
\begin{split}
	&E= -8 \lambda_1^2 -4 \ell_1 y \lambda_1 \lambda_2 - y^2\left(\frac12(\ell_2^2+\ell_1^2) +4 \ell_2 y +8 y^2\right) \lambda_2^2 \, ,\\ 
	&L=  \ell_1 \frac{\lambda_1}{\lambda_2} + \frac14(\ell_2^2 + \ell_1^2)y + \ell_2 y^2\, \, ,\qquad r=-2\kappa=2\,.
\end{split}
\end{equation}

Given the holographic interpretation of the present Ansatz, that we mentioned above, we expect that it contains solutions descending from the  AdS$_6$ solutions of \cite{DHoker:2016ujz,dhoker-gutperle-uhlemann}. 


\subsubsection{Another sub-Ansatz} 
\label{ssub:another}

Another possibility we can explore is 
\begin{equation}
	L= L(y)\, ,\qquad  V= V(y)\,.
\end{equation}
The third equation in (\ref{eq:sep-total}) is then automatically satisfied. The first two imply that $E$ is a polynomial of total degree 2 in $\lambda_i$: 
\begin{equation}\label{eq:Eanother}
	E= \sum_{0\le a+b\le 2} L_{ab} \lambda_1^a \lambda_2^b \ .
\end{equation}
Moreover, they determine $L_{20}$ and $L_{02}$ in terms of $L$ and $V$. The fourth, quadratic equation in (\ref{eq:sep-total}) then gives a system of six ODEs in the $y$ coordinate, one for each monomial $\lambda_1^a \lambda_2^b$, $0\le a+ b \le 2$. 

One observes that the system simplifies substantially by assuming $L_{11}=0$. Moreover, the ODE corresponding to the monomials of total degree $<2$ are linear in $L_{ab}$, $a+b=0,1$ once the ODEs corresponding to total degree 2 have been solved. The latter are now equations for $L$ and $V$, and can be solved, for example, with a power-law assumption. This way we get
\begin{equation}
\begin{split}
	&\kappa=0 \, ,\qquad V= r y^2 \, ,\qquad L= \ell - r y^4 \, ,\qquad L_{20}=L_{02}=0 \, ,\\ 
	&L_{10}= \ell_1 L \,, \qquad L_{01}= \ell_2 L \, ,\qquad L_{00}= L \left(\frac{\ell_2^2}{4y} + \ell_0 - \frac{\ell_1^2}{12} y^3 \right) \,, 	
\end{split}
\end{equation} 
where $r$, $\ell$, $\ell_a$, $a=0,1,2$ are constants.


\section*{Acknowledgements}
We would like to thank John Estes for useful correspondence. AP and AT are grateful to the Mainz Institute for Theoretical Physics (MITP) for its hospitality during the completion of this work. Our research was supported in part by INFN and by the European Research Council under the European Union's Seventh Framework Program (FP/2007-2013) -- ERC Grant Agreement n. 307286 (XD-STRING). AT was also supported by the MIUR-FIRB grant RBFR10QS5J ``String Theory and Fundamental Interactions''. AP is also supported by the Knut and Alice Wallenberg Foundation under grant Dnr KAW 2015.0083.

\appendix

\section{GL$(2,\mathbb{R})$ transformation}
\label{GL2R}
In this appendix we show how $c^{IJ}$ can be set equal to $2 \delta^{IJ}$, by a GL$(2,\mathbb{R})$ transformation of $\eta^I_{i+}$.

First, by rescaling the $\chi$'s in \eqref{spin_decomp} we can set
\begin{equation}
c^{11} = c^{22} = 2 \ .
\end{equation}
Our analysis then splits into two cases: (a) $|c^{12}| \neq 2$ and (b) $|c^{12}| = 2$. 
For case (a) we define  $x \equiv \frac{1}{2} c^{12}$, so that $x^2\neq 1$. Then the GL$(2,\mathbb{R})$ map 
\begin{equation}
\left(\begin{array}{c} 
\eta^1_{i+} \\ \eta^2_{i+}
\end{array}\right) 
\rightarrow  
\left(
\begin{array}{cc}-1 & 0 \\ 
-\frac{x}{\sqrt{1-x^2}} &
 \frac{1}{\sqrt{1-x^2}}
\end{array}
\right)
\left(\begin{array}{c} 
\eta^1_{i+}  \\ \eta^2_{i+} 
\end{array}\right)
\end{equation}
leaves the norms $\|\eta_{i+}^I\| = e^A$ invariant and in the new basis 
\begin{equation}
\overline{\eta^{(1}_{i+}} \eta^{2)}_{i+} = \R (\overline{\eta^1_{i+}} \eta^2_{i+}) = 
\frac{1}{2} c^{12} e^A = 0 \ .
\end{equation}
In the second case $\overline{\eta^{(1}_{i+}} \eta^{2)}_{i+} = \R (\overline{\eta^1_{i+}} \eta^2_{i+}) = \frac{1}{2}c^{12}e^A = \pm e^A$, and from the Cauchy--Schwarz inequality
\begin{equation}
\sqrt{\R(\overline{\eta^1_{i+}}\eta^2_{i+})^2 + \I(\overline{\eta^1_{i+}}\eta^2_{i+})^2 } = |\overline{\eta^1_{i+}}\eta^2_{i+}| \leq \sqrt{\|\eta^1_{i+}\|\|\eta^2_{i+}\|} = e^A
\end{equation}
it follows that $\I( \overline{\eta^1_{i+}}\eta^2_{i+} ) =0$; in addition since the inequality is saturated $\eta^1_i$ and $\eta_i^2$ should be proportional. The factor of proportionality is fixed by their norms and inner product $|\overline{\eta^1_{i+}}\eta^2_{i+}| = e^A$
to be $\pm 1$. But in this case there is only $\mathcal{N} = 1$ supersymmetry, as can be readily inferred from the 10$d$ spinor decomposition Ansatz.

\section{Spinors and $\mathcal{G}$-structures}

We look at the $\mathcal{G}$-structures defined by spinors in $1+3$ and $6$ dimensions. They are characterized by a set of tensors constructed as spinor bilinears which we will assemble into bispinors $\epsilon\bar{\epsilon}$, since the latter, via the Fierz expansion (schematically)\footnote{$\g_{m_1 \dots m_p}$ denotes the antisymmetric product of $\g_{m_1}, \dots, \g_{m_p}$.}
\begin{equation}
\epsilon\bar{\epsilon} \propto \sum_p \frac{1}{p!} \gamma^{m_p\dots m_1} \bar{\epsilon} \gamma_{m_1 \dots m_p} \epsilon \ ,
\end{equation}
and the map
\begin{equation}
\gamma^{m_p\dots m_1} \to dx^{m_p} \wedge \dots \wedge dx^{m_1}
\end{equation}
can be treated as polyforms.

\subsection{Dimension $d=1+3$}\label{spin(1,3)}

In this appendix we examine the identity structure defined by two spinors, $\zeta^1_+$ and $\zeta^2_+$, of positive chirality in $1+3$ dimensions\footnote{One chiral spinor, $\zeta_+$, defines an $\mathbb{R}^2$ structure; see for example\cite[Sec. 4.1.1]{10d}.}. 

The generators of Cliff$(1, 3)$ satisfy $\{\g^\alpha , \g^\beta\} = 2 \eta^{\alpha\beta}$, $\alpha, \beta = 0,1,2,3$, where $\eta^{\alpha\beta}$ is the Minkowski metric of ``mostly plus'' signature; they are chosen so that $(\g^\alpha)^\dagger = \g^0 \g^\alpha \g^0$. The chirality operator is $\g_5 \equiv  -i \g^0 \g^1 \g^2 \g^3$ and has the property $(\g^5)^2 = \mathbb{I}$. We introduce the intertwiner $B$ that relates $\g^\alpha$, $\alpha = 0,1,2,3$ and its complex conjugate $(\g^\alpha)^*$ as $\g^\alpha B =  B (\g^\alpha)^*$. It satisfies $B^* =  B^{-1}$ and $B^\dagger = B^{-1}$. The complex conjugate of a spinor $\zeta$ is then $\zeta^c \equiv B \zeta^*$; note that $\zeta^{cc} =  \zeta$. Complex conjugation changes chirality. 

We look at the case of a ``strict'' identity structure where $\zeta^1_+$ and $\zeta_+^2$ are orthogonal i.e. $(\zeta^1_+)^\dagger \zeta_+^2 = 0$. 
Employing the Fierz identity and
\begin{equation}
\g_{\alpha_1 \dots \alpha_k} = 
(-1)^{\frac{k(k-1)}{2}} \frac{-i}{(4-k)!} \epsilon_{\alpha_1 \dots \alpha_4} \g^{\alpha_{k+1}\dots\alpha_4} \g_5 \ ,
\end{equation}
the following expansions for the bispinors can be obtained:\footnote{$\e^{i_1i_2 \dots i_n}$ denotes the wedge product $\e^{i_1} \wedge \e^{i_2} \wedge \dots \e^{i_n}$.}
\begin{align}
\zeta^1_+ \overline{\zeta^1_+} &= \frac{1}{4}(\e^+ + i * \e^+) \ , \qquad
\zeta^1_+ \overline{\zeta^1_-} = \frac{1}{4} \e^{+1} \ , \\ \nn
\zeta^2_+ \overline{\zeta^2_+} &= \frac{1}{4}(\e^- + i * \e^-) \ , \qquad
\zeta^2_+ \overline{\zeta^2_-} = -\frac{1}{4} \e^{-\bar 1} \ , 
\end{align}
Here $\overline \zeta \equiv \zeta^\dagger \gamma^0$ and $\zeta_- = (\zeta_+)^c$. The set of 1-forms
$\{ \e^+, \e^-, \e^1, \e^{\bar 1} \}$ make up a complex frame defining the identity structure. A real frame 
can be constructed as
\begin{equation}
\e^0 = \frac{1}{2}(\e^+ + \e^-) \ , \qquad 
\e^3 = \frac{1}{2}(\e^+ - \e^-) \ , \qquad 
\e^1 = \frac{1}{2} (\e^1 + \e^{\bar 1}) \ , 
\qquad \e^2 = -\frac{i}{2} (\e^1 - \e^{\bar 1}) \ .
\end{equation}
The volume element is
$\vol_4 = \e^{0123}$ and the Hodge star is defined via $a \wedge * b = (a,b) \vol_4$, where $(.,.)$ is the inner product with respect to the Minkowski metric. Thus, for example,
\begin{equation}
* 1 = \e^{0123} \ , \qquad * \e^0 = - \e^{123} , \qquad * \e^3 = - \e^{012}. 
\end{equation}
Furthermore, 
\begin{equation}
\zeta^1_+ \overline{\zeta^2_+} = \frac{1}{4}(\e^1 + i * \e^1) \ , \qquad
\zeta^1_+ \overline{\zeta^2_-} = -\frac{1}{4}(1 + \tfrac{1}{2}\e^{-+} + \tfrac{1}{2}\e^{1\bar 1} - i * 1) \ .
\end{equation}

We record the following identities
\begin{equation}
\zeta^I_- \overline{\zeta^J_-} = B (\zeta^I_+ \overline{\zeta^J_+})^* B^{-1} \ , \qquad
\zeta^I_- \overline{\zeta^J_+} = B (\zeta^I_+ \overline{\zeta^J_-})^* B^{-1} \ . 
\end{equation}
for generic spinors $\zeta$'s. Bearing in mind that $B^{-1}\g^\alpha B =  (\g^\alpha)^*$, we conclude that,  a plus to minus interchange is equivalent to complex conjugation. For example
\begin{equation}
\zeta^1_- \overline{\zeta^1_-} = \frac{1}{4}(\e^+ - i * \e^+) \ .
\end{equation}
Finally,
\begin{equation}
\zeta^I_+ \overline{\zeta^J_{\mp}} = - (-1)^{\frac{k(k+1)}{2}} \zeta^J_{\pm} \overline{\zeta^I_-} \ .
\end{equation}

\subsection{Dimension $d=6$}\label{spin(6)}

In this appendix we take a look at the $\mathcal{G}$-structures defined by chiral spinors in six dimensions. Given a representation $\{ \g_1, \g_2, \dots, \g_6 \}$ of Cliff(6) we introduce
\begin{equation}
\mathfrak{g}_1 \equiv \frac{1}{2}(\g_1 + i \g_2) \ , \qquad
\mathfrak{g}_2 \equiv \frac{1}{2}(\g_3 + i \g_4) \ , \qquad
\mathfrak{g}_3 \equiv \frac{1}{2}(\g_5 + i \g_6) \ . 
\end{equation}
The Cliffora algebra then takes the form
\begin{equation}
\{\mathfrak{g}_{\frak a}, \mathfrak{g}_{\bar {\frak  b}} \} = \delta_{{\frak a} \bar {\frak b}} \ , \qquad
\{\mathfrak{g}_{\frak a}, \mathfrak{g}_{\frak b} \} 
= \{\mathfrak{g}_{\bar {\frak a}}, \mathfrak{g}_{\bar {\frak b}} \} = 0 \ , \qquad 
{\frak a}, {\frak b} = 1,2,3 \ ,
\end{equation}
where $\mathfrak{g}_{\bar 1} = \tfrac{1}{2}(\mathfrak{g}_1 - i \mathfrak{g}_2)$ etc.

We take $\ket{\downarrow\downarrow\downarrow}$ as the state which is annihilated by all $\mathfrak{g}_{\frak a}$. Starting from $\ket{\downarrow\downarrow\downarrow}$ and acting with $\mathfrak{g}_{\bar{\frak a}}$ we can construct the $2^3$-dimensional Dirac representation of Spin(6). We denote $\ket{\uparrow\downarrow\downarrow} = \mathfrak{g}_{\bar{1}} \ket{\downarrow\downarrow\downarrow}$ etc. Expanding $\gamma_7 \equiv i \g_1 \dots \g_6$ in terms of $\mathfrak{g}_{\frak a}$, $\mathfrak{g}_{\bar {\frak a}}$, we conclude that spinors with an even number of $\uparrow$ have positive chirality while spinors with an odd number of $\uparrow$ have negative chirality. The intertwiner $B$, which relates $\g_a$, $a=1,2,\dots,6$ and $(\g_a)^*$ as $\g_a B = - B (\g_a)^*$,  interchanges $\downarrow$ and $\uparrow$ and hence chirality. For example $B \ket{\downarrow\downarrow\downarrow} = \ket{\uparrow\uparrow\uparrow}$.

A chiral spinor $\eta_+ \equiv \ket{\downarrow\downarrow\downarrow}$ defines an SU(3) structure, characterized by a real 2-form $J$ and a decomposable complex 3-form $\Omega$, as
\begin{equation}
\eta_+\overline{\eta_+} = \frac{1}{8}\left(1 - iJ - *J + i *1 \right) \ , \qquad
\eta_+\overline{\eta_-} = -\frac{1}{8} \Omega \ ,
\end{equation}
where $\overline{\eta_+} \equiv \eta_+^\dagger$,
\begin{equation}
-iJ = \frac{1}{2} (\e^{1 \bar 1} + \e^{2 \bar 2} + \e^{3 \bar 3}) \ , \qquad
\Omega = \e^{1 2 3} \ ,
\end{equation}
and $\{\e^1$, $\e^2$, $\e^3\}$ are a complex frame.
$J$ obeys 
\begin{equation}
J \wedge J \wedge J = 6 \vol_6 \ , \qquad *J = \frac{1}{2} J \wedge J \ .
\end{equation}
Accordingly,
\begin{equation}
\eta_+\overline{\eta_+} = \frac{1}{8} e^{-iJ} \ .
\end{equation}

Two chiral spinors $\eta^1_+$ and $\eta^2_+$ define an SU(2) structure as follows: we take $\eta^1_+ \equiv \ket{\downarrow\downarrow\downarrow}$ and $\eta^2_+$ to be orthogonal. The stabilizer group $\mathcal{G}$ of $\eta^1_+$ in Spin(6) $\simeq$ SU(4) is SU(3). We can thus perform an SU(3) transformation that leaves $\eta^1_+$ invariant and sets $\eta^2_+ = \ket{\uparrow\uparrow\downarrow} = \mathfrak{g}_3 \ket{\uparrow\uparrow\uparrow}$. Then 
\begin{equation}
\e^3 \ , \qquad 
\omega \equiv \iota_{\e^{\bar 3}} \Omega \ , \qquad 
-i j \equiv - i J - \frac{1}{2} \e^{3 \bar 3} \ ,
\end{equation}
define an SU(2) structure in six dimensions, where $\e^3$ is the 1-form bilinear constructed out of $\eta^1_+$ and $\eta^2_+$.

Along the same lines four chiral spinors $\ket{\downarrow\downarrow\downarrow}$, $\mathfrak{g}_1 \ket{\uparrow\uparrow\uparrow}$, $\mathfrak{g}_2 \ket{\uparrow\uparrow\uparrow}$ and $\mathfrak{g}_3 \ket{\uparrow\uparrow\uparrow}$ define a (strict) identity structure.

We record the following identities
\begin{equation}
\eta^I_- \overline{\eta^J_-} =  B (\eta^I_+ \overline{\eta^J_+})^* B^{-1} \ , \qquad
\eta^I_- \overline{\eta^J_+} = B (\eta^I_+ \overline{\eta^J_-})^* B^{-1} \ . 
\end{equation}
Bearing in mind that $B^{-1}\g_a B =- \g^*_a$, we conclude that,  in the first case a plus to minus interchange is equivalent to complex conjugation but in the second case minus complex conjugation. 
Finally,
\begin{equation}
\eta^I_+ \overline{\eta^J_{\mp}} =  (-1)^{\frac{k(k+1)}{2}} \eta^J_{\pm} \overline{\eta^I_-} \ .
\end{equation}

\section{Coordinates on the $\{v^5,v^6\}$-subspace}
\label{v5v6coord}

We want to show here that \eqref{dv56} implies \eqref{eq:v56dx12}. Let us call
\begin{equation}\label{eq:wvv}
	\beta \equiv v^5 + i v^6 \ , \qquad \alpha \equiv 2v - i v^1 \  ,
\end{equation}
so that \eqref{dv56} reads
\begin{equation}\label{eq:dw3aw3}
	d \beta = \alpha \wedge \beta \ .
\end{equation}

Separating \eqref{dv56} in real and imaginary parts, we see that it reads 
\begin{equation}\label{eq:aab}
	dv^a = a^{ab} \wedge v^b \ ,\qquad a^{56}+ i a^{66}=i(a^{55}+ia^{65})=i \alpha \ .
\end{equation}
By the dual version of the Frobenius theorem (see for example \cite[Th.~B.3.2]{wald}), it follows that the four-dimensional distribution $D_4 \subset T$ orthogonal to $v^5$ and $v^6$, $D_4 = \{ X \in T | \iota_X v^a=0,\, a=5,6\}$, is integrable. This means that $D_4$ is a foliation: there exist (generically) four-dimensional leaves, such that the union of all of them is the whole manifold $M_6$. In other words, at every point there is a leaf that goes through that point. These can be parameterized by $6-4=2$ real numbers, which we can call $x_1$, $x_2$, so that the leaves can be labeled as $L_{x_1,x_2}$. We can also use $x_1$, $x_2$ as coordinates on $M_6$. They are constant on each leaf, since they parameterize them. In other words $dx_i \perp D_4$, $i=1,2$. But by definition the $v^a$ are also orthogonal to each leaf. So we conclude 
\begin{equation}\label{eq:vdx}
	v^a= m^{ai} dx_i \ .
\end{equation}
We can also define the differential $d_L$ such that $d= d_L + dx_i \wedge \partial_{x_i}$. 

Using the coordinates we have just introduced, we can write $\beta= \beta^i dx_i$; moreover, we can decompose $\alpha=\alpha_L + \alpha^i dx_i$. From \eqref{eq:dw3aw3} it now follows that
	\begin{equation}
		d_L \beta^i = \alpha_L \beta^i \ .
	\end{equation} 
From this we conclude that $d_L\left(\beta^2/\beta^1\right)=0$. So $w \equiv \beta^2/\beta^1$ is a function of $x_1$ and $x_2$ only. Now $\beta = \beta^1 (dx_1 + w dx_2)$; but $dx_1 + w dx_2$ defines an almost complex structure in two dimensions, which is always integrable. Thus there exists a complex coordinate $z$ such that $(dx_1 + w dx_2)$ is proportional to $dz$. Hence 
	\begin{equation}\label{eq:wdz}
		\beta = e^\varphi dz \ .
	\end{equation}
Redefining $x_1$, $x_2$ so that $z=x_1 + i x_2$, we arrive at \eqref{eq:v56dx12} with $\varphi = 2\Sigma - i \psi$.

\bibliography{at}
\bibliographystyle{utphys}

\end{document}